# Soft point-contact Andreev reflection spectroscopy in a palm-type cubic anvil-pressure cell

Qingxin Dong [1,2], Fengrui Shi [3], Yan Zhang [3], Tong Shi [1,4], Yi Liu [3], Shaoheng Ruan [1,2], Zhongjin Wu [1,2], Jianping Sun [1,2], Zhaoming Tian [4], Yoshiya Uwatoko [6,7], Guanghan Cao [3], Xin Lu [3], Bosen Wang（王铂森）[1,2]* and Jin-Guang Cheng [1,2]*

[1] Beijing National Laboratory for Condensed Matter Physics and Institute of Physics, Chinese Academy of Sciences, Beijing 100190, China

[2] School of Physical Sciences, University of Chinese Academy of Sciences, Beijing 100049, China

[3] School of Physics, Zhejiang University, Hangzhou 310058, China

[4] Wuhan National High Magnetic Field Center and School of Physics, Huazhong University of Science and Technology, Wuhan 430074, China

[5] School of Physics and Optoelectronic Engineering, Shandong University of Technology, 266 Xincun West Road, Zibo, 255000, P. R. China

[6] Institute for Solid State Physics, University of Tokyo, Kashiwanoha 5-1-5, Kashiwa, Chiba 277-8581, Japan

[7] Faculty of Science and Engineering, Tokyo City University, Tamazutsumi 1-28-1, Setagaya-ku, Tokyo 158-8557, Japan

*Corresponding authors: bswang@iphy.ac.cn (BSW); jgcheng@iphy.ac.cn (JGC)

## Abstract

We have implemented soft point-contact Andreev reflection spectroscopy (PCARS) in a palm-type cubic-anvil pressure cell by combining a substrate anchoring strategy with an external wire-splitting technique. This design enables the stable formation of multiple point-contact junctions under hydrostatic pressures up to 15 GPa. Benchmark measurements on the elemental superconductor Nb demonstrate high reproducibility and yield a zero-temperature superconducting gap $\Delta(0) \approx 1.3$ meV with a ratio $2\Delta(0)/k_B T_c \approx 3.3$. We further apply this technique to the kagome metal superconductor $CsCr_3Sb_5$ and the bilayer nickelate superconductor $La_2PrNi_2O_7$. Pronounced zero-bias conductance peaks are observed, and their evolution with temperature, magnetic field and applied pressure is investigated, together with the superconducting gap magnitude and possible pairing symmetries. These measurements provide spectroscopic evidence consistent with unconventional superconductivity in these materials. Our work establishes a robust experimental platform that bridges macroscopic electrical transport and microscopic spectroscopic probes, opening a new avenue for investigating pairing symmetry in a wide range of pressure-induced unconventional superconductors.

## Introduction

Understanding the microscopic pairing mechanism in pressure-induced superconductors remains a central challenge in condensed matter physics, largely owing to the scarcity of reliable probes that can access microscopic superconducting-state properties under high pressure. Soft point-contact Andreev reflection spectroscopy (PCARS) is a particularly powerful technique for this purpose [1-4]: it directly measures the differential conductance of a normal-metal-superconductor interface, thereby revealing the density of states and key features of the superconducting gap function. Importantly, the soft-contact geometry greatly facilitates integration into high-pressure setups, offering the possibility to continuously track the evolution of the superconducting order parameter with pressure and to gain deeper insight into the underlying pairing mechanisms [5-11].

Initial attempts to perform PCARS at high pressure have relied on diamond anvil pressure cells (DACs) and piston-cylinder cells (PCCs) [5-11]. Although these efforts have yielded valuable information, significant limitations remain. The inherently constricted sample space in DACs (typically ≤ 100 μm) poses severe challenges for the stable creation of nanoscale junctions and the results can be difficult to interpret unambiguously. Furthermore, the pronounced pressure inhomogeneity introduced by the solid pressure-transmitting media commonly used in DACs compromises the reliability of measurements on many unconventional superconductors that are highly sensitive to non-hydrostatic stresses. While PCCs offer a sufficiently large sample volume, their moderate pressure range (< 3 GPa) restricts broader application. Consequently, implementing PCARS in a high-pressure cell that simultaneously provides a hydrostatic environment and a large sample volume has become an urgent technical need.

The cubic-anvil cell (CAC) has emerged as a powerful apparatus for investigating novel physical properties of quantum materials under high pressures [12-14], and has played a particularly important role in the discovery of pressure-induced unconventional superconductors [15-20]. As shown in Fig. 1(a), its unique design combines triaxial symmetric compression and a liquid pressure-transmitting medium, ensuring an isotropic and hydrostatic pressure environment. Together with a sample volume that is several orders of magnitude larger than that of a DAC, the CAC can generate extreme pressures exceeding 15 GPa while maintaining excellent pressure homogeneity [15-20]. The self-clamped construction also facilitates integration with ultra-low temperatures (down to the millikelvin) and high magnetic fields (up to 12 T) [21]. This distinctive combination of extreme conditions has made the CAC a powerful tool to discover and identify unconventional superconductivity in strongly correlated electronic systems [15-20, 22, 23]. Despite these advantages, soft PCARS has not yet been implemented in a CAC because of the complexities associated with its anvil geometry and electrical connections.

In this work, we realize high-pressure PCARS measurements inside a CAC by combining a substrate anchoring strategy with an external wire-splitting technique. Our experimental design achieves a nearly 100% success rate and has been applied to several representative superconductors, including the elemental superconductor Nb, the newly discovered kagome metal superconductor $CsCr_3Sb_5$ and the bilayer nickelate superconductor $La_2PrNi_2O_7$. Analysis of the PCARS spectra yields the magnitude of the superconducting gap and its evolution with temperature, magnetic field, and physical pressure, all of which are consistent with the previously reported electrical transport and electronic phase diagrams. At the same time, the measurements provide preliminary information on superconducting pairing symmetry in these materials. By bridging macroscopic electrical transport and microscopic superconducting gap, the established PCARS platform in the present work deepens the understanding of superconducting mechanisms and offers a powerful new route for probing pairing symmetry in a wide range of pressure-induced unconventional superconductors.

## Experimental Methods

## Design challenges and solutions

Integrating PCARS into a CAC presents two major challenges. The first is the mechanical stability of the point-contact junction. Conventionally, the junctions are formed by applying a silver-paste dot less than 50 μm in diameter, but the weak adhesive strength at the interface leaves the junction vulnerable to the pulling and twisting of gold wires during the sample transfer into the pressure cell and subsequent application of high-pressure loading. To address this issue, we adopt a substrate anchoring strategy. As shown in Figs. 1(b) and 1(c), the sample is mounted on an insulating substrate (e.g., $Al_2O_3$ or $SiO_2$) slightly larger than the sample, and the gold wires forming the junction are securely fixed to this substrate. This configuration isolates the delicate junction from tensile stresses induced by wire manipulation during transfer or pressure loading.

The second challenge is the limited number of electrical channels in a standard CAC. As shown in Fig. 1(a), the internal geometry comprises six anvils, with the four side anvils serving both to apply compressive force and to provide primary electrical pathways to the sample. In a standard configuration, each electrode connects to a single anvil, which suffices for traditional four-probe transport measurements but severely restricts PCARS measurements: only a single point-contact junction can be probed per pressure cycle, substantially reducing experimental throughput. To overcome this bottleneck, we implement an external wire-splitting technique. Each primary feed-through wire is branched outside the anvil body to create multiple independent measurement circuits, as illustrated in Fig. 1(b). Although this introduces additional series resistance from the extended gold leads and the tungsten carbide (WC) anvils, the contribution is negligible (< 0.1 Ω) compared with the typical junction resistance (several to tens of ohms) and can be disregarded. The modification therefore does not impair the intrinsic quality of the PCARS spectra while dramatically improving measurement efficiency.

## Measurement principle of PCARS

The core of a PCARS measurement is the acquisition of the differential conductance $dI/dV$ of a normal-metal-superconductor (NS) junction as a function of the applied DC bias voltage $V$. We employ a standard modulation technique in which a small-amplitude AC excitation is superimposed on a DC bias current. If the DC current is $I$ and the AC modulation is $\delta I \sin(\omega t)$, the total current through the junction is $I + \delta I \sin(\omega t) = I_{NS}(V + \delta v \sin(\omega t))$. The differential conductance signal at bias voltage $V_0$ is obtained after the Taylor expansion:

$$I_{NS}(V+\delta v\sin(\omega t)) = I(V_0) + \left.\frac{dI}{dV}\right|_{V_0}\delta v\sin(\omega t) + \left.\frac{d^2I}{2dV^2}\right|_{V_0}(\delta v\sin(\omega t))^2$$

$$= I(V_0) + \left.\frac{dI}{dV}\right|_{V_0}\delta v\sin(\omega t) + \left.\frac{d^2I}{4dV^2}\right|_{V_0}\delta v^2[1-cos\,(2\omega t)]$$

By analyzing the junction's response, the first derivative of the current-voltage characteristic $dI/dV$ is directly encoded in the amplitude of the AC voltage signal at the modulation frequency $\omega$.

Experimentally, a nearly constant AC current source and a well-defined small-signal excitation $\delta I$ are realized by passing the AC voltage output of a lock-in amplifier (Stanford Research Systems SR830) through a large series resistor (1 MΩ). A DC current source (Keithley 6221) supplies the bias current $I$ and the DC bias voltage $V$ across the junction is measured with a nanovoltmeter (Keithley 2182A), as depicted in Fig. 1(c). The lock-in amplifier simultaneously measures the resulting AC voltage $\delta V$ across the junction. Since the excitation is a well-defined small-signal current $\delta I$, the differential conductance is obtained as $dI/dV \approx \delta I/\delta V$. Sweeping the DC current $I$ while recording $\delta V$ produces the $dI/dV(V)$ spectrum, which directly reflects the electronic density of states of the superconductor.

## Results and Discussion

### Point-contact Andreev reflection spectroscopy on the elemental superconductor Nb

To validate the experimental setup, we first performed PCARS measurements on the elemental superconductor Nb under hydrostatic pressure in the CAC. Figure 2(a) shows the temperature-dependent electrical resistance of Nb at 2.0 GPa, revealing a superconducting transition temperature $T_c = 9.1$ K, consistent with previous reports [24, 25]. The corresponding PCARS spectrum at ambient pressure (AP) and 1.8 K (outside the CAC) is shown in Fig. 2(b); a pair of pronounced coherence peaks appears at approximately ±1.3 meV on both sides of zero bias, characteristic of conventional isotropic s-wave pairing [26]. When the pressure is increased to 2 GPa in the CAC [Fig. 2(c)], the overall spectral shape remains unchanged, confirming the reliability of the PCARS measurement under pressure. Notably, in both cases, in addition to the superconducting gap features, obvious dips in $dI/dV$ are observed at low temperatures at energies just above the gap (around ± 5 meV at 2 K); the bottom energy of these dips decreases with increasing temperature. Such dip features are commonly observed in point-contact spectroscopy and have been attributed to critical-current effects or local heating when the junction deviates from the pure ballistic limit.

Compared with the AP spectrum in Fig. 2(b), the superconducting gap obtained at 2 GPa shows a slight reduction, Fig. 2(c). Meanwhile, the flanking dip features become more pronounced, possibly indicating that pressure reduces the ballistic contribution to the junction conductance. With increasing temperature, the coherence peaks gradually weaken and merge, accompanied by an overall suppression of the spectroscopic signal, reflecting the thermal depairing of the superconducting state. Near 10 K, above $T_c$, the PCARS signature vanishes completely, confirming that the sample has entered the normal state. Fitting the spectra with an isotropic *s*-wave model yields the temperature evolution of the superconducting gap $\Delta(T)$, as plotted in Fig. 2(d). The extracted zero-temperature gap $\Delta(0) \approx 1.3$ meV and the ratio $2\Delta(0)/k_BT_c \approx 3.3$ are in excellent agreement with the established values [24, 25], demonstrating that the CAC-based PCARS platform faithfully reproduces known superconducting gap parameters.

**Point-contact Andreev reflection spectroscopy on the kagome superconductor $CsCr_3Sb_5$**

We next applied PCARS to the newly discovered chromium-based kagome superconductor $CsCr_3Sb_5$ [19]. At AP, the compound exhibits long-range antiferromagnetic order below $T_N \approx 55$ K accompanied by the formation of density waves, Fig. 3(a) and Ref. [19]. Under hydrostatic pressures of 4-8 GPa, superconductivity emerges at low temperatures with $T_c$ between 2.5 and 6 K. The pressure-temperature phase diagram reveals non-Fermi-liquid behavior and signatures of magnetic quantum criticality, suggesting an unconventional pairing mechanism [15, 16, 27].

To probe directly its superconducting properties directly, we performed PCARS measurements on $CsCr_3Sb_5$ in the CAC under pressure. The spectrum at 5 GPa [Fig. 3(b)] displays a sharp zero-bias conductance peak (ZBCP), in stark contrast to the double-peak structure of conventional BCS superconductors [26]. Such a ZBCP is widely regarded as a hallmark of unconventional superconductivity, typically arising from sign-changing order parameters that support Andreev bound states at interfaces. In kagome metals, theoretical studies have predicted that strong antiferromagnetic fluctuations and flat-band-enhanced correlations can stabilize nodal pairing states, including the $d_{xy}$ and $d_{x^2-y^2}$-wave pairing symmetries [27].

As temperature increases, the ZBCP progressively broadens and decreases in amplitude, eventually merging into a smooth background above $T_c$. Figure 3(c) shows the normalized spectra, where the superconducting contribution is clearly resolved after removing the normal-state background. Fits using a phenomenological *d*-wave BTK model capture the essential spectral features. The extracted superconducting gap $\Delta$ and broadening factor $\Gamma$ are plotted in Fig. 3(d), yielding $\Delta(0) \approx 2.3$ meV and a strong-coupling ratio $2\Delta(0)/k_BT_c \approx 8.2$. These results provide spectroscopic evidence consistent with unconventional pairing, likely *d*-wave character, in pressurized $CsCr_3Sb_5$, in good agreement with recent experiments and theoretical predictions [27-30].

**Point-contact Andreev reflection spectroscopy on the polycrystalline nickelate superconductor $La_2PrNi_2O_7$**

We further extended PCARS measurements to higher pressures, focusing on the recently discovered nickelate high-$T_c$ superconductor $La_2PrNi_2O_7$ [18, 31]. As shown in Fig. 4(a), its resistivity at 7 GPa exhibits semiconducting behavior over the entire measured temperature range. At 11.5 GPa, a slight low-temperature downturn in resistivity appears, suggesting the possible onset of superconductivity [18]. However, the corresponding PCARS spectra at this pressure show no clear superconducting signature, Fig. 4(b), which may be attributed to a small superconducting volume fraction or pressure inhomogeneity caused by solidification of the pressure-transmitting medium. These observations are consistent with previous reports [18].

Upon increasing the pressure to 13.5 GPa, a pronounced resistive drop sets in below approximately 68 K, confirming the emergence of a superconducting state [18]. At this pressure, well-defined Andreev reflection spectra are obtained [Fig. 4(c)], displaying a sharp ZBCP that is widely regarded as strong evidence for unconventional superconducting pairing [5,7]. Temperature-dependent PCARS measurements show that the high-bias background gradually flattens upon warming, while a discernible hump near zero bias persists up to about 80 K, above which the spectra become indistinguishable from the normal-state background. This behavior is fully consistent with the resistance transition and marks the complete suppression of superconductivity [18].

To better resolve the superconducting features, we normalized the spectra by subtracting a smooth background. The normalized spectra, together with fits using a nodal *d*-wave model, are shown in Fig. 4(d). The fits capture the ZBCP reasonably well, although small deviations remain, possibly arising from multiple gap components or inelastic scattering effects [1]. Furthermore, we should note that it is difficult to distinguish between the *d*- and *s*+- waves in PCARS measurements, as there is no unified wave function expression for *s*+- wave. However, phenomenologically, if the s± gap has nodes, the corresponding PCARS spectrum may also exhibit a sharp zero-bias conductance peak similar to that expected for a d-wave gap. From the fits, we extracted the temperature dependence of the superconducting gap Δ, as plotted in Fig. 4(e). The gap opens just below $T_c$ and increases rapidly upon cooling, reaching a zero-temperature value of $\Delta(0) \approx 6$ meV, which is close to the reported value of the small gap in previous studies [5, 6, 32, 33]. Additional gap-like features appear at higher bias voltages, but they are relatively broad and ill-defined and show a strong dependence on junction quality.

In previous studies, the measured superconducting gap of $La_3Ni_2O_7$ varies slightly with sample quality and pressurization method, and some superconducting gaps are even unobservable, with its PCARS showing one gap in one junction while two or three gaps in another junction [5, 6, 33]. For example, PCARS studies of single-crystal $La_3Ni_2O_7$ in the DAC have revealed two *s*-wave gaps structure at higher pressure of 23 GPa, with a larger $\Delta_{s1}(0) \approx 23$ meV and a smaller $\Delta_{s2}(0) \approx 6$ meV [5]. A three-gap structure of $La_3Ni_2O_7$ single crystal was also reported at 20 GPa in the DAC, comprising a *d*-wave gap with a magnitude of about 9 meV, and two other *s*-wave gaps at ≈ 16 and 26 meV [6]. These different superconducting gap features in PCARS may be related to the orientation of single crystal or pressure environment. This discrepancy may also reflect the random grain orientations in polycrystalline samples and possible non-hydrostatic pressure effects. At the same time, such discrepancies may also arise from the fact that the pressure of 13.5 GPa in the present study is insufficient to realize a bulk superconducting state [18] and the observed spectra in Fig. 4(c) are likely from the filamentary superconducting phase. Future PCARS measurements on $La_2PrNi_2O_7$ at pressures ~ 20 GPa are required to fully capture its superconducting gap structure. Nevertheless, the clear observation of a ZBCP in this and other pressurized superconductors provides strong evidence that unconventional pairing mechanisms are prevalent among pressure-induced superconductors near magnetic quantum critical points [15-17, 19, 20].

The PCARS technique developed in this work offers distinct advantages: it can yield reliable estimates of superconducting gaps magnitudes and allow one to infer possible superconducting pairing symmetries, thereby providing microscopic evidence for superconductivity that is complementary to electrical transport measurements. However, the method also presents significant limitations. Because it primarily probes the local electronic structures, PCARS is susceptible to extrinsic signals and may produce misleading results arising from contact resistance issues; notably, contact resistance can generate spurious signals through voltage division effects, necessitating extremely careful data analysis. Looking forward, the CAC-based PCARS platform requires further improvement. Key objectives include providing a large sample space and a pressure environment with excellent hydrostaticity, while also addressing the lead resistances inevitably introduced when incorporating multiple electrodes to improve experimental efficiency. In addition, optimizing contact preparation and refining signal discrimination will be crucial to enhancing both the measurement throughput and the intrinsic reliability.

## Conclusion

In summary, we have implemented soft point-contact Andreev reflection spectroscopy in a cubic-anvil cell by combining a substrate anchoring strategy with an external wire-splitting technique. Measurements on the conventional BCS superconductor Nb confirm the reliability and stability of this method. Applications to the unconventional superconductors $CsCr_3Sb_5$ and $La_2PrNi_2O_7$ at high pressures reveal zero-bias conductance peaks that provide spectroscopic evidence consistent with unconventional pairing mechanisms. This approach extends high-pressure studies from macroscopic transport and magnetic susceptibility to the microscopic level of the superconducting gap structure and pairing symmetry, opening new avenues for investigating pressure-induced unconventional superconductivity.

## Acknowledgements

We thank Dr. Ningning Wang and Dr. Gang Wang for preparing high-quality polycrystalline $La_2PrNi_2O_7$ samples. This work was supported by the National Key R&D Program of China (Grants No. 2023YFA1406100, No. 2023YFA1607400, No. 2024YFA1408400, No. 2022YFA1403800, and No. 2022YFA1403203), the National Natural Science Foundation of China (Grant No. 12025408, 12074414, and U23A6003), JSPS KAKENHI Grant Number 25K00951, the CAS President's International Fellowship Initiative (2024PG0003), and the Outstanding member of Youth Promotion Association of CAS (Y2022004), the CAS Superconducting Research Project (grant no. SCZX-0101). This work was carried out at the cubic anvil cell (CAC) station of the Synergetic Extreme Condition User Facility (SECUF, https://cstr.cn/31123.02.SECUF)

## Competing interests

The authors declare no competing interests.

## Data Availability

The data supporting the findings of this study are available from the corresponding author upon reasonable request. Source data are provided with this paper.

## Figure Captions

**Figure 1** (a) The cubic-anvil pressure apparatus (CAC) and sample configuration within the MgO gasket. (b) Schematic of the external wire-splitting technique and photograph of a point-contact junction prepared for Andreev reflection measurements. (c) Schematic of the measurement circuit for point-contact spectroscopy and resistance measurements.

**Figure 2** (a) Temperature-dependent resistance of the elemental superconductor Nb at 2 GPa. (b) Point-contact Andreev reflection spectrum at ambient pressure at 1.8 K. (c) Point-contact Andreev reflection spectra at 2 GPa and various temperatures. Solid lines are fits using an isotropic *s*-wave model. (d) Temperature dependence of the superconducting gap extracted from the fits.

**Figure 3** (a) Normalized resistance $R/R_{298K}$ of $CsCr_3Sb_5$ at 0 GPa, and 5 GPa, respectively. (b) Point-contact spectra at 5 GPa at various temperatures. (c) Normalized point-contact spectra at different temperatures, with fits using a *d*-wave model. (d) Temperature dependence of the superconducting gap $\Delta$ and broadening factor $\Gamma$.

**Figure 4** (a) Temperature-dependent resistance of $La_2PrNi_2O_7$ at various pressures. (b) Point-contact Andreev reflection spectrum at 11.5 GPa at 4.2 K. (c) Point-contact Andreev reflection spectra at 13.5 GPa and various temperatures. (d) Normalized $dI/dV$ spectra with fits using a *d*-wave model. (e) Temperature dependence of the superconducting gap; the red dashed line represents a BCS fit.

# Figure 1

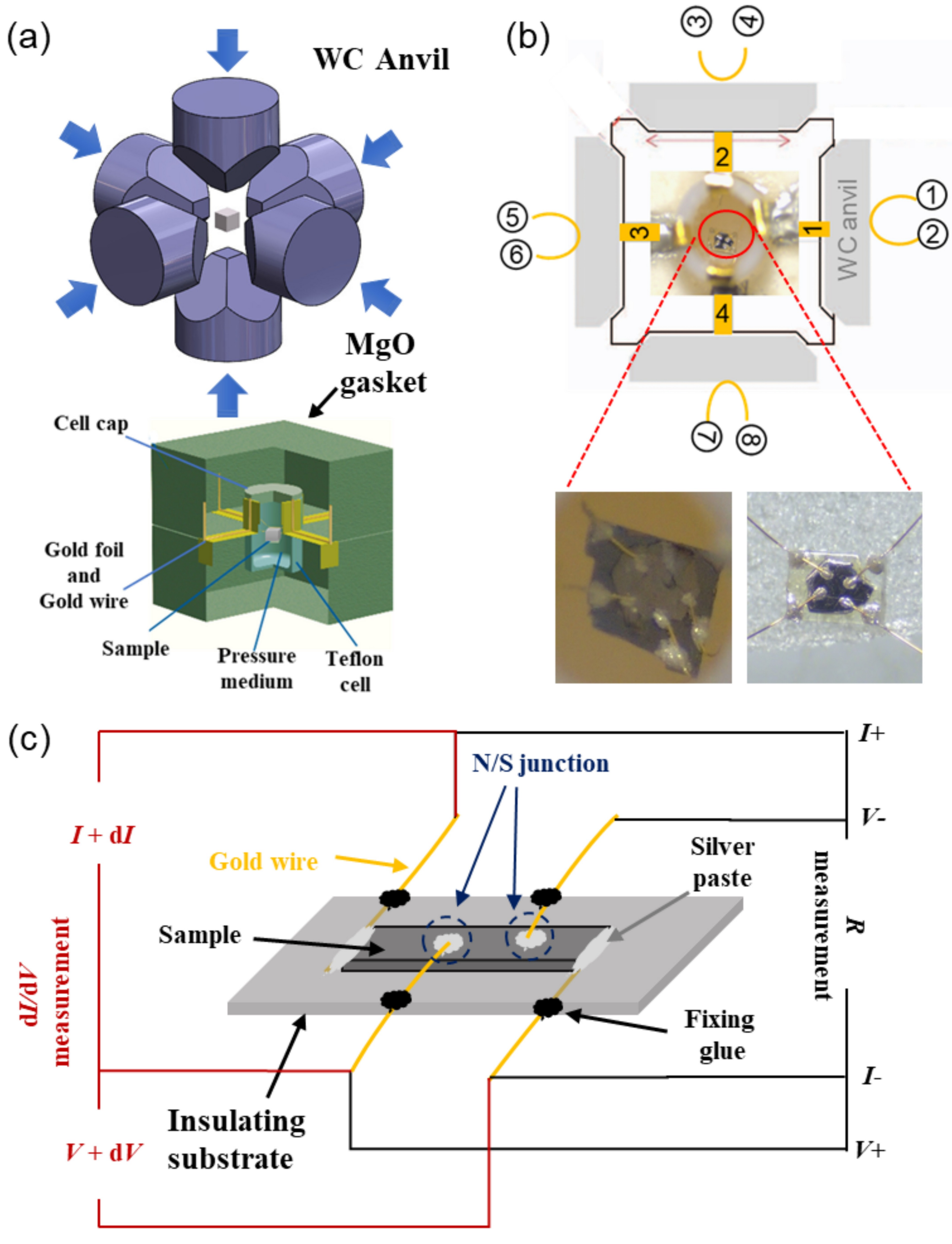

(a)
WC Anvil
MgO
gasket
Cell cap
Gold foil
and
Gold wire
Sample
Pressure
medium
Teflon
cell
(b)
WC anvil
(c)
N/S junction
I+
V-
I + dI
Gold wire
Silver
paste
Sample
dI/dV
measurement
R
measurement
Fixing
glue
Insulating
substrate
I-
V + dV
V+

# Figure 2

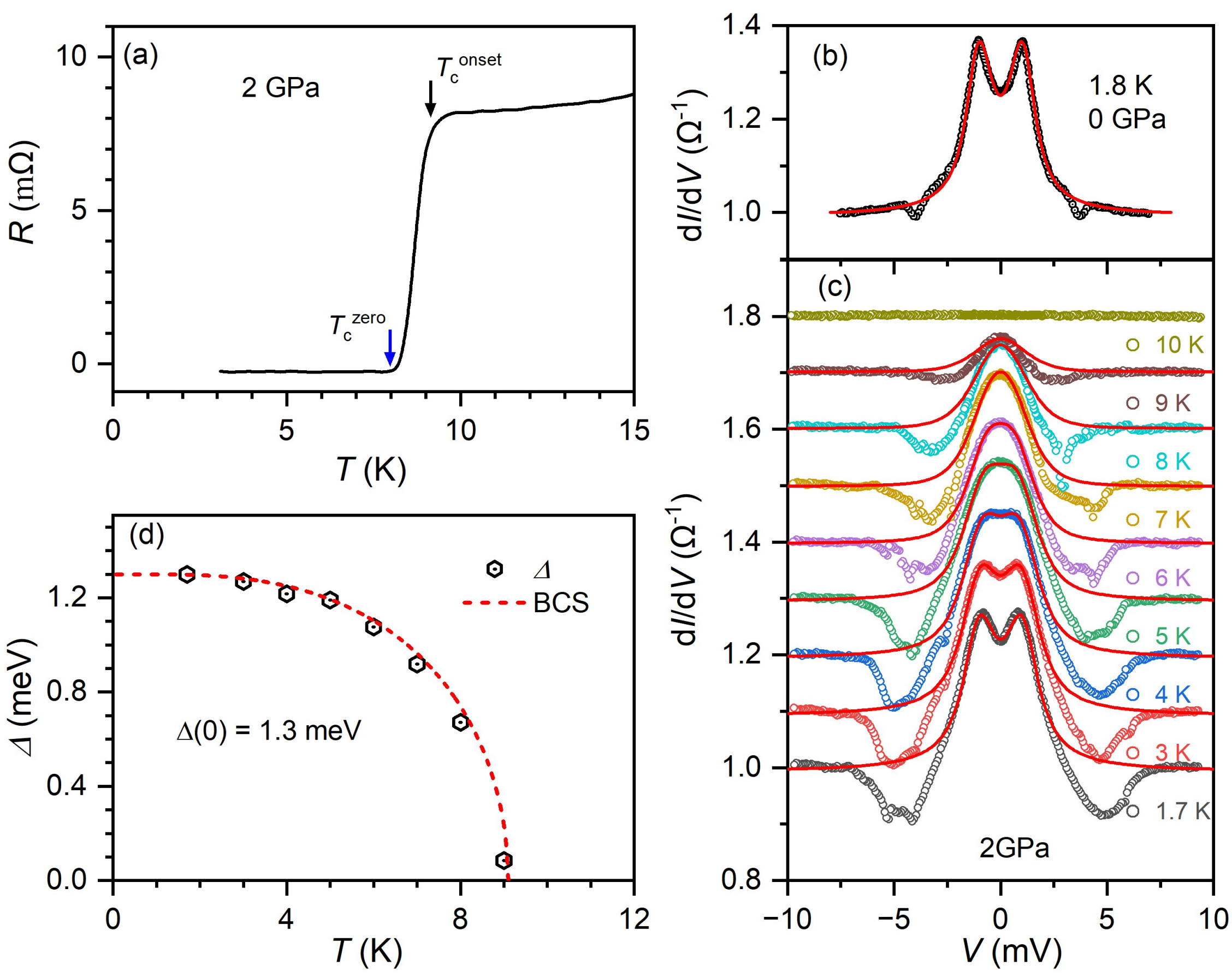

(a)
2 GPa
T_c^onset
T_c^zero
R (mΩ)
T (K)
(b)
1.8 K
0 GPa
dI/dV (Ω^-1)
(c)
10 K
9 K
8 K
7 K
6 K
5 K
4 K
3 K
1.7 K
2GPa
dI/dV (Ω^-1)
V (mV)
(d)
Δ
BCS
Δ(0) = 1.3 meV
Δ (meV)
T (K)

**Figure 3**

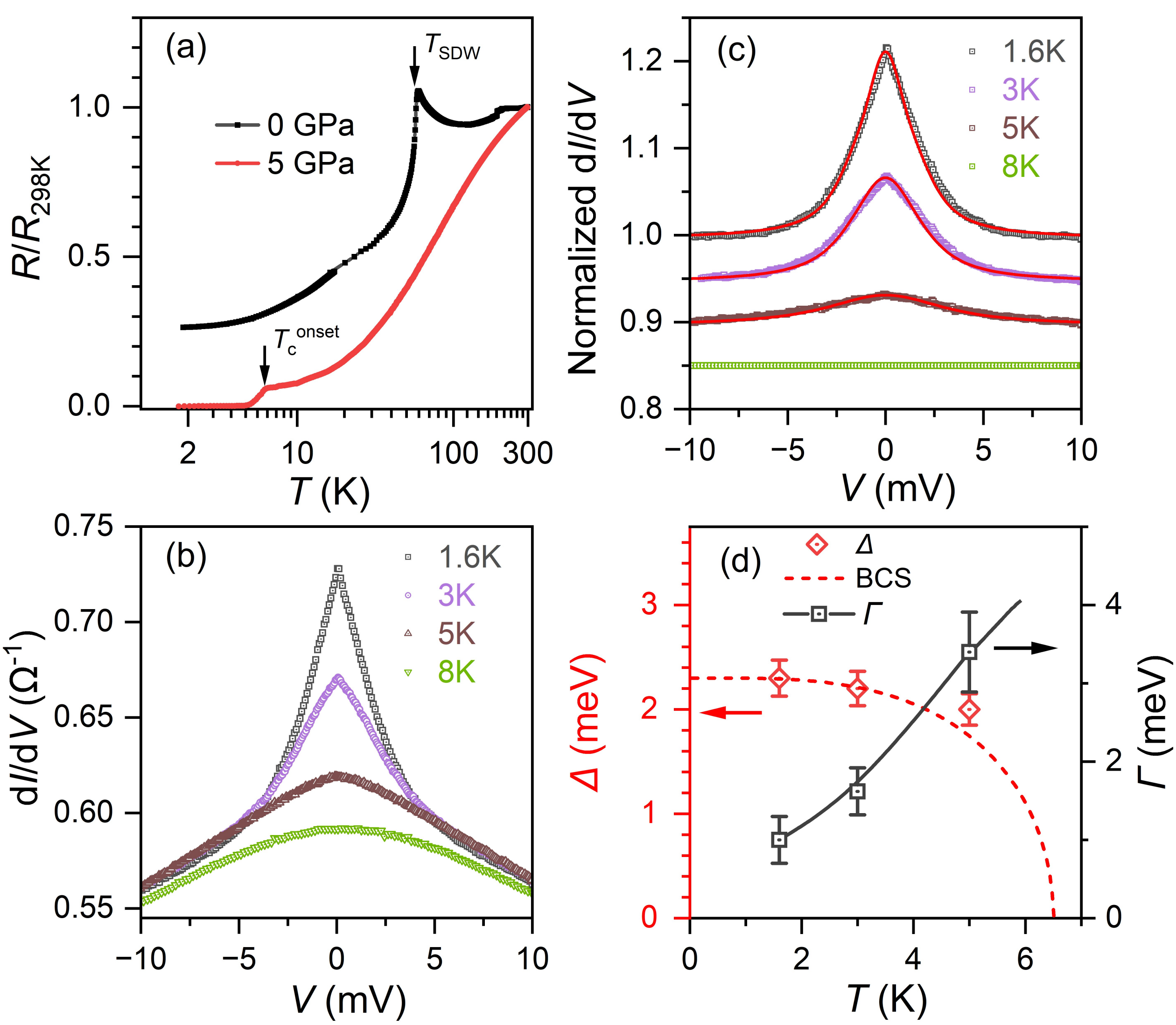

(a)
$T_{SDW}$
0 GPa
5 GPa
$T_c^{onset}$
$R/R_{298K}$
$T$ (K)
(b)
1.6K
3K
5K
8K
d$I$/d$V$ ($\Omega^{-1}$)
$V$ (mV)
(c)
1.6K
3K
5K
8K
Normalized d$I$/d$V$
$V$ (mV)
(d)
$\Delta$
BCS
$\Gamma$
$\Delta$ (meV)
$\Gamma$ (meV)
$T$ (K)

**Figure 4**

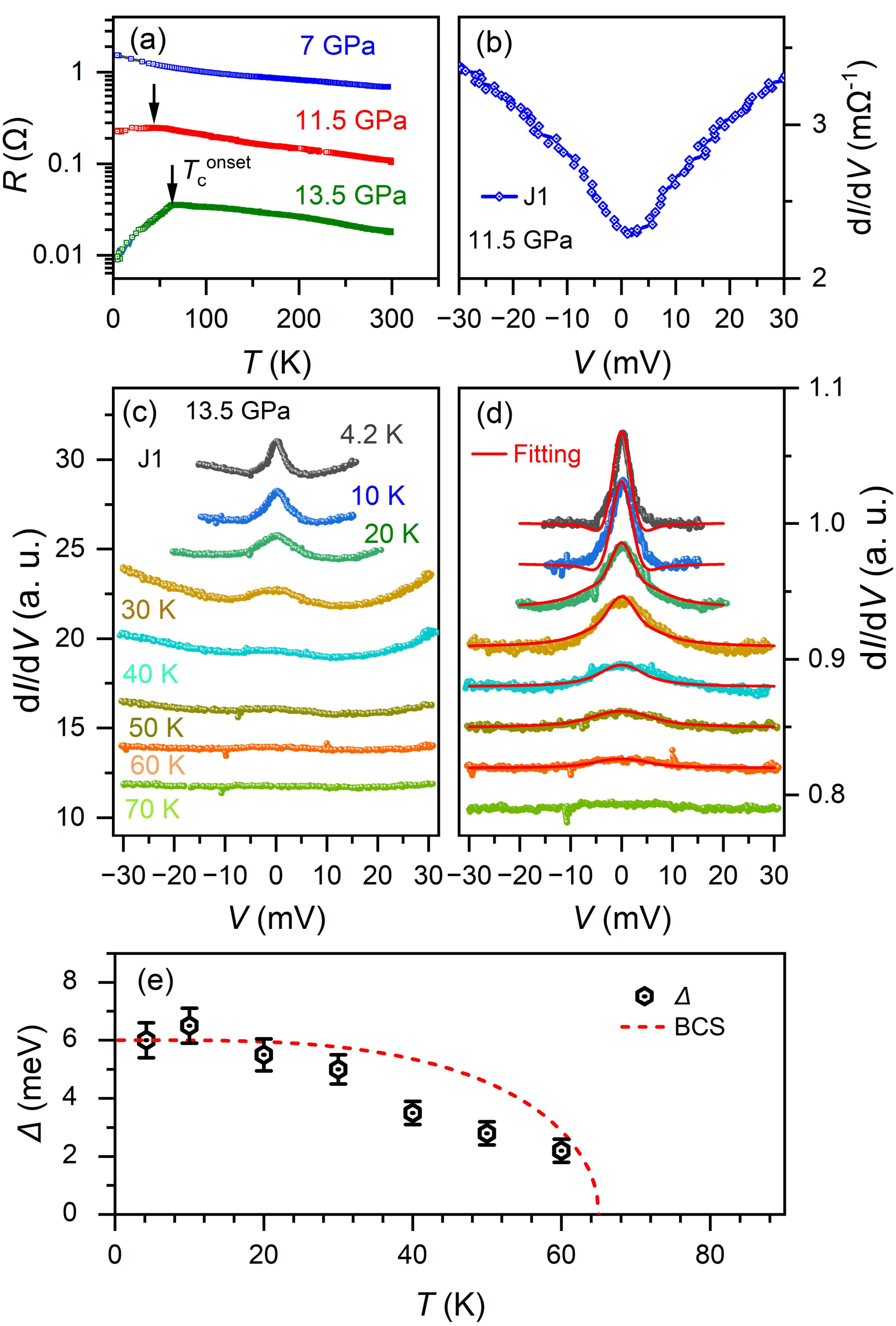

(a)
7 GPa
11.5 GPa
13.5 GPa
$T_c^{onset}$
R (Ω)
T (K)
(b)
J1
11.5 GPa
dI/dV (mΩ⁻¹)
V (mV)
(c)
13.5 GPa
J1
4.2 K
10 K
20 K
30 K
40 K
50 K
60 K
70 K
dI/dV (a. u.)
V (mV)
(d)
Fitting
dI/dV (a. u.)
V (mV)
(e)
Δ
BCS
Δ (meV)
T (K)